\documentclass[9pt,twocolumn,twoside]{osajnl}

\journal{ol} 

\setboolean{shortarticle}{true}

\title{Integrated Random Projection and Dimensionality Reduction by Propagating Light in Photonic Lattices}

\author[1,2,*]{Mohammad-Ali Miri}

\affil[1]{Department of Physics, Queens College of the City University of New York, Queens, New York 11367 USA}
\affil[2]{Physics Program, The Graduate Center, City University of New York, New York, New York 10016, USA}

\affil[*]{mmiri@qc.cuny.edu}

\begin{abstract}

It is proposed that the propagation of light in disordered photonic lattices can be harnessed as a random projection that preserves distances between a set of projected vectors. This mapping is enabled by the complex evolution matrix of a photonic lattice with diagonal disorder, which turns out to be a random complex Gaussian matrix. Thus, by collecting the output light from a subset of the waveguide channels, one can perform an embedding from a higher-dimension to a lower-dimension space that respects the Johnson-Lindenstrauss lemma and nearly preserves the Euclidean distances. It is discussed that distance-preserving random projection through photonic lattices requires intermediate disorder levels that allow diffusive spreading of light from a single channel excitation, as opposed to strong disorder which initiates the localization regime. The proposed scheme can be utilized as a simple and powerful integrated dimension reduction stage that can greatly reduce the burden of a subsequent neural computing stage.

\end{abstract}

\begin{document}

\maketitle

Interesting properties such as energy-efficiency and the possibility of long-range interactions make photonics appealing for unconventional computing \cite{wetzstein2020inference}. Although a universal classical optical computer seems to be a far reach, there is a great interest in efficient optical implementation of certain computing tasks. Photonics particularly holds a great promise for implementing high-speed and energy-efficient linear computing operations. In addition, strong nonlinear effects suggest opportunities for applications such as hybrid or all-optical optimization and machine learning. To this end, apart from technological limitations, a major challenge in optical computing is a lack of methods and algorithms that are compatible with the photonics hardware. Subsequently, there is a need for identifying and devising computing methods and algorithms that can be efficiently implemented with the existing photonics hardware.

\begin{figure}
    \centering
    \includegraphics[width=1\linewidth]{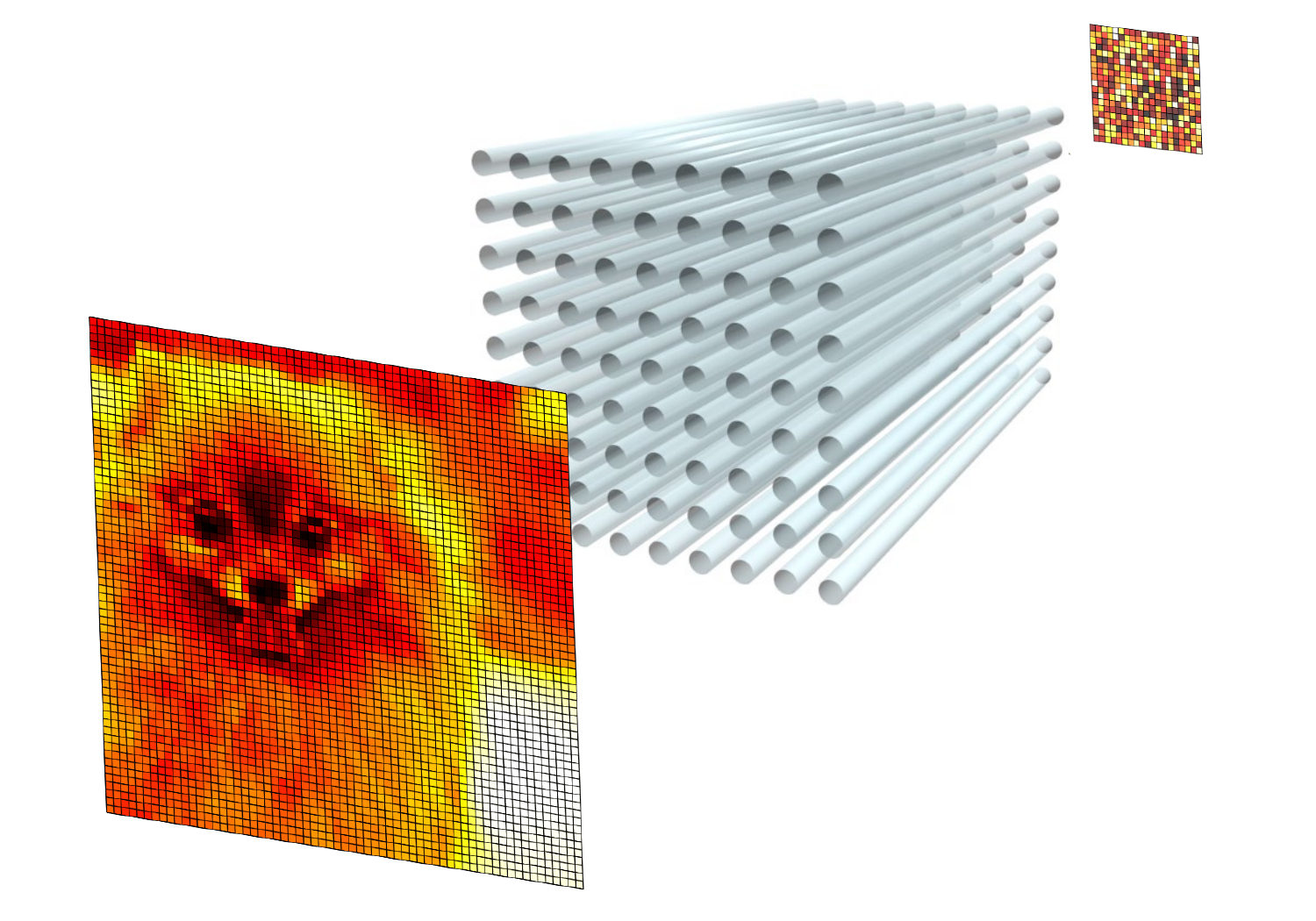}
    \caption{A schematic of random projection by propagating a spatially-modulated wavefront through a photonic lattice and evaluating the transmitted wavefront at a fewer number of output ports.}
    \label{fig1}
\end{figure}

Here, a new method for optical implementation of dimensionality reduction through random projection is proposed. Random projection is an efficient approach for dimensionality reduction in very-high-dimensional datasets \cite{bingham2001random}. In this technique, data are linearly mapped from an original high-dimensional space to a low-dimensional subspace, while nearly preserving their Euclidean distances during this transformation. Thus, the transformed data points remain distinguishable, while their reduced dimensions greatly simplifies a desired processing task such as classification. In random projection, this dimension reduction is done simply through multiplication by a random matrix with certain statistical properties \cite{bingham2001random}.

Recently, it is shown that random projection can be performed optically by propagating spatially modulated light through a strongly scattering medium \cite{liutkus2014imaging, saade2016random}. What enables such an optical realization is the direct analogy of multiple scattering in random media with a linear random mapping between the coherent light wavefront at the input and output planes. In fact, it is known that a partial description of such a scattering medium is modeled by a random complex matrix with entries that are drawn independently from a Gaussian distribution \cite{vellekoop2015feedback, Gigan_2017, goetschy2013filtering, davy2015transmission}.

In this letter, it is shown that optical random projection can be performed by propagating light in photonic lattices with transverse disorder, and thus without a need to a scattering medium. This realization allows for integrated photonic implementation of random projection. In the proposed setting, data are injected as complex modal amplitudes of the electric field in a waveguide array, while collected from a fewer number of waveguide channels in the output to perform dimension reduction (Fig.~\ref{fig1}). It is shown that a proper implementation of random projection requires diagonal disorder of a critical strength that ensures diffusive transport of light rather than ballistic or localized transport.

\begin{figure}[t]
    \centering
    \includegraphics[width=1.0\linewidth]{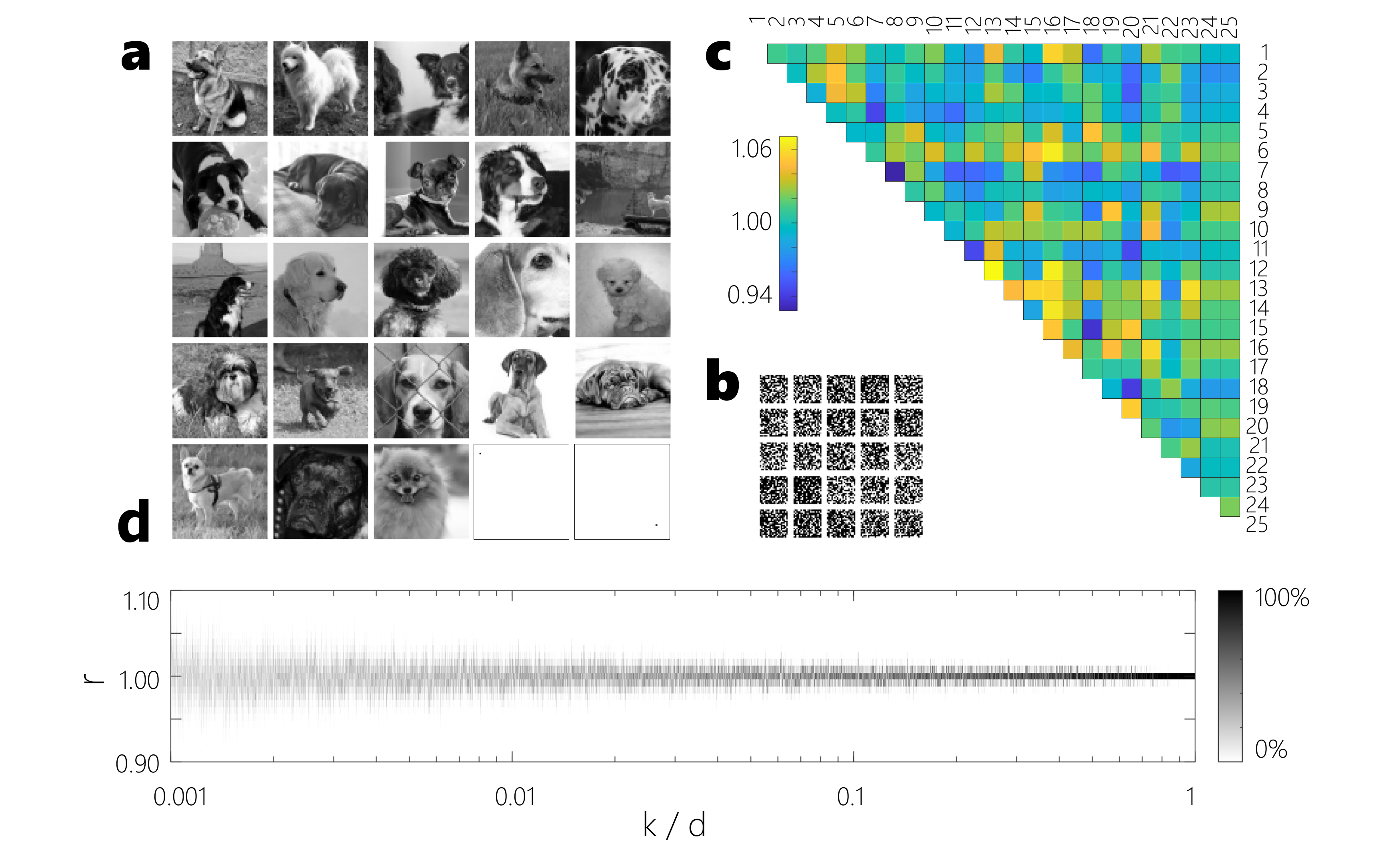}
    \caption{Random projection of an exemplary set of 25 vectors for $d=4096 \rightarrow k=400$. (a) A set of $64\times 64$ pixel grayscale images from the downsampled ImageNet dataset \cite{ImageNetLink} along with two binary images for exploring extreme cases. (c) The ratio of the pairwise distances between data points in the transformed and original spaces, $ \sqrt{d/k} \|T\mathbf{a}_i-T\mathbf{a}_j\|/\|\mathbf{a}_i-\mathbf{a}_j\|$. (d) The distribution of the pairwise distance ratio $r$ for different dimensional reduction factors $k/d$. Here, a 2D array of $d = 64 \times 64$ waveguides with uniform nearest-neighbor coupling $\kappa_0$ is considered, while $\kappa_0 L = 100$, when $L$ is the array length, and the diagonal detunings $\delta_i$ are drawn from a uniform random distribution, i.e., $\delta_i \sim \mathcal{U}(-\delta_0,\delta_0)$, when $\delta_0 / \kappa_0 = 0.5$.}
    \label{fig2}
\end{figure}


The goal of random projection is to perform a linear transformation $\mathbf{x} \mapsto \mathbf{y}$ of a set of $n$ data points $\mathbf{x}_1,\cdots,\mathbf{x}_n$ in $d$-dimensional space to $\mathbf{y}_1,\cdots,\mathbf{y}_n$ in a $k$-dimensional subspace ($k \ll d$), such that, ideally, $\|\mathbf{y}_i-\mathbf{y}_j\| \approx \|\mathbf{x}_i-\mathbf{x}_j\|$ for all $(i,j)$ pairs of the data points. The core of such a projection is the Johnson–Lindenstrauss (JL) lemma \cite{johnson1984extensions}, which guarantees the existence of a linear transformation, $f:\mathbb{R}^d \rightarrow \mathbb{R}^k$, that preserves Euclidean distances with a tolerance $\varepsilon$, i.e.,
\begin{equation}
    (1-\varepsilon) \| \mathbf{x}_i-\mathbf{x}_j \|^2 \leq  \| f(\mathbf{x}_i)-f(\mathbf{x}_j) \|^2 \leq (1+\varepsilon) \| \mathbf{x}_i-\mathbf{x}_j \|^2.
\end{equation}
where, $\| \cdot \|^2$ represent the L$^2$ norm. For a given number of data points $n$, the reduced dimension $k$ should be in the order of $k_0 = \mathcal{O}(\varepsilon^{-2} \ln{n})$ to maintain all distances within this range \cite{johnson1984extensions}. The original work of Johnson and Lindenstrauss built the proof on matrices with their rows selected from random orthogonal matrices \cite{johnson1984extensions}. The proof was later extended to random Gaussian matrices \cite{frankl1988johnson}, and to sub-Gaussian matrices (tail of the distribution falls below the Gaussian tail) \cite{matouvsek2008variants}. In addition, it was shown that random matrices with binary entries $t_{ij} \in \{-1,+1\}$, with equal probabilities of $\pm 1$, or with trinary entries $t_{ij} \in \{-1, 0, +1\}$, while statistically two-thirds of the elements are zero, can respect the JL property, thus serve as random projectors \cite{achlioptas2003database}. The motivation for the latter work was the efficient computational implementation of random projection for high-dimensional data. However, the aforementioned trinary matrix is still a dense matrix, and subsequently there has been a great interest in alternative matrices that allow for fast random projections \cite{ailon2009fast}.


It is straightforward to show that the JL lemma can be extended to the complex domain and for complex-valued mappings, i.e., when the input and output vectors and the projector matrix are in general complex. Thus, the idea becomes relevant to linear optical devices that deal with the complex phasor of the electric field. However, it is important to identify a proper optical system that can perform a projection that respects the JL lemma. As discussed in the following, such a projection can be realized effectively by propagating light through a optical waveguide array, which under proper transverse disorder simulates an orthogonal complex Gaussian matrix.



Here, a photonic lattice composed of an array of waveguides with evanescent nearest-neighbor coupling is considered as depicted schematically in Fig.~\ref{fig1}. The propagation of light in photonic lattices in the presence of disorder in the propagation constants (diagonal) and coupling coefficients (off-diagonal) has been intensely investigated in the past \cite{segev2013anderson}. In particular, the transition of light propagation from ballistic to diffusive and localization regimes have been experimentally demonstrated in one- \cite{lahini2008anderson} and two-dimensional \cite{schwartz2007transport} optical waveguide arrays with diagonal and off-diagonal disorder \cite{di2013einstein}. Considering an array of $d$ waveguides, the evolution of the complex modal amplitudes $\mathbf{u}(z) = (u_1(z), \cdots, u_d(z))^t$ along the propagation distance $z$ is governed by \cite{christodoulides2003discretizing}:
\begin{equation}
    \frac{d}{dz} \mathbf{u} = i H \mathbf{u}
\end{equation}
where, $H$ is a real-symmetric matrix with diagonal elements $h_{ii}=\delta_i$ representing detuning of the waveguide propagation constants from a reference value and off-diagonal $h_{ij} = h_{ji} = \kappa_{ij}$ representing coupling constants that are non-zero for nearest neighbors and zero otherwise. Considering a finite length $L$ for the waveguide array, the complex field amplitudes at the input plane $z=0$, are mapped to those at $z=L$ according to $\mathbf{u}(z=L) = U \mathbf{u}(z=0)$, where $U=e^{i H L}$. From the available $d$ output ports, one can collect the light from a sample of $k$ channels. In this manner, one can effectively implement a linear mapping $\mathbb{C}^{d} \rightarrow \mathbb{C}^{k}$:
\begin{equation}
    \mathbf{b} = T \mathbf{a}
\end{equation}
where, $\mathbf{a} \in \mathbb{C}^{d}$ is the vector of the complex field amplitudes in the input $\mathbf{u}(z=0)$, $\mathbf{b} \in \mathbb{C}^{k}$ is a subset of the complex field amplitudes at the output $\mathbf{u}(z=L)$, and the transfer matrix $T \in \mathbb{C}^{k\times d}$ is a row-sampled sub-matrix of the evolution operator $U = e^{i H L}$.

Clearly, the evolution matrix $U$ is a unitary matrix, i.e., $U^{\dagger} U = U U^{\dagger} = I_{d \times d}$, where $I_{d \times d}$ is the $d \times d$ identity matrix. Thus, the transfer matrix $T$, which is utilized for random projection has orthogonal rows, $T T^{\dagger} = I_{k \times k}$. While clearly the columns of $T$ are not orthogonal, ideally it is desired to make those close to orthogonal, i.e., $T^{\dagger} T \approx I_{d \times d}$, which is expected to happen by randomness of the matrix elements. In the following, first, random projection is explored numerically by considering diagonal disorder of proper strength. Next, the criteria for the suitable type of disorder is discussed.

To numerically exemplify random projection, a square lattice array of $d$ waveguides with diagonal disorder is considered, while the length of the waveguide array is chosen to ensure that a single-channel excitation fills the entire array. The original data vectors are fed into the $d$ channels and the output of a sample of $k$ ports ($k \leq d$) is evaluated as the projected vector of reduced dimensions. A given real-valued data point can be encoded in the phase of the input vector $\mathbf{a}$ which generally results in a complex projected output vector $\mathbf{b}$. In practice, the complex vector $\mathbf{b}$ can be utilized as a low-dimensional representation of the original data $\mathbf{a}$ to be fed directly to a subsequent optical processing device, which could be an optical neural network. It is worth noting that in the proposed scheme, the power of the projected vectors is reduced by an approximate factor of $k/d$. However, this factor is nearly uniform for all projected vectors, thus the distances remain nearly preserved by simply scaling the projected vectors by a factor of $\sqrt{k/d}$.

Figures~\ref{fig2}(a,b) show a set of images with $d=4096$ pixels and the corresponding projected vectors with $k=400$ pixels. Here, a set of grayscale $64 \times 64$ pixel images from the downsampled ImageNet dataset \cite{ImageNetLink} is considered along with two binary images with an extreme distance in the original space. As clearly depicted in Fig.~\ref{fig2}(c), the pairwise distances between the projected vectors is preserved within a small range centered at unity. The pairwise distances are evaluated for different dimension reduction factors $k/d$ an plotted as histograms in Fig. \ref{fig2}(d), which as expected shows an increase in the standard deviation from unity as the $k/d$ decreases.

\begin{figure}[t]
    \centering
    \includegraphics[width=1\linewidth]{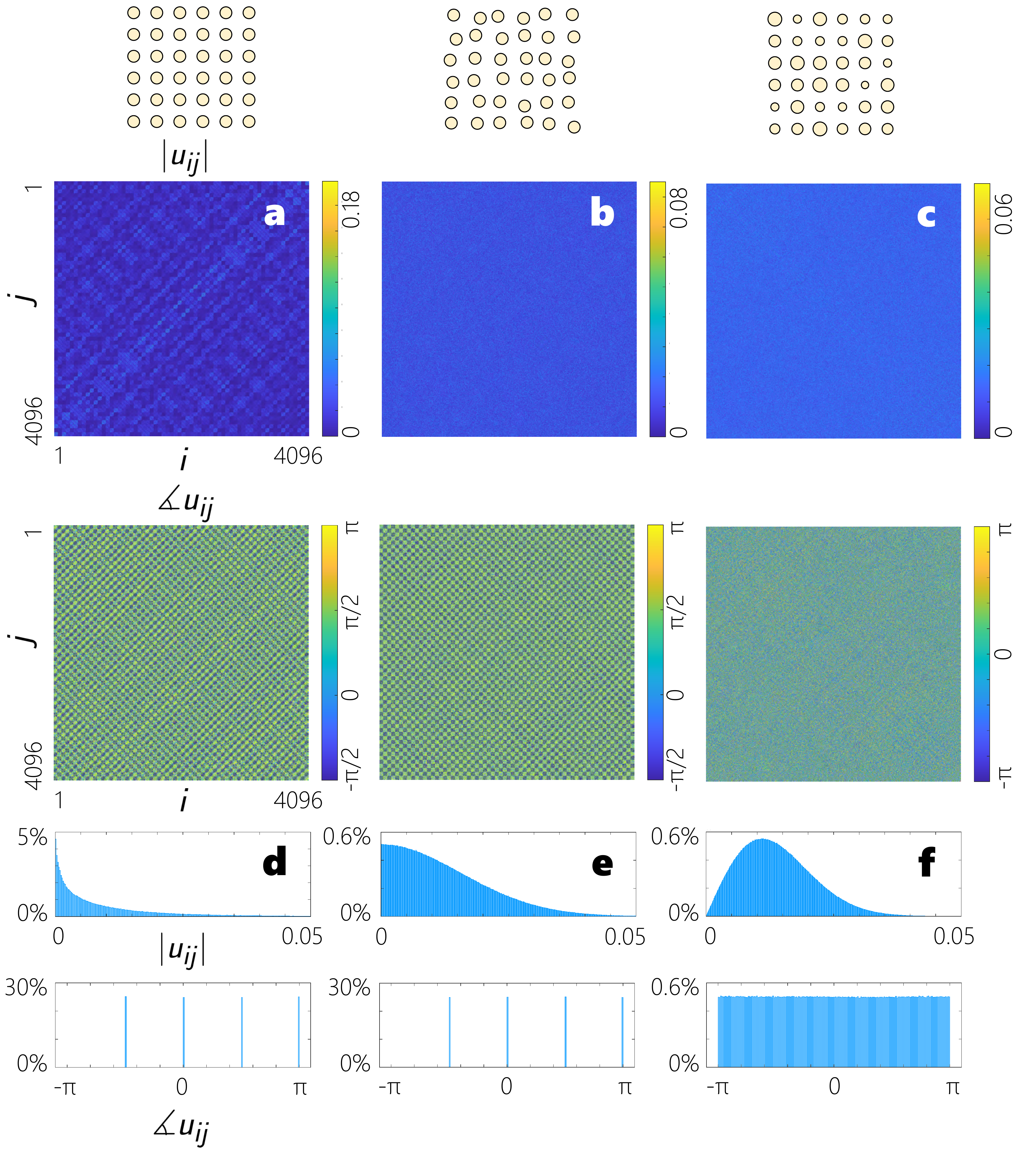}
    \caption{The entries of the complete transfer matrix $U$ for a photonic lattice composed of a rectangular array of waveguides with nearest-neighbor coupling, (a) without disorder, (b) with off-diagonal disorder, (c) with diagonal disorder. In (a), all waveguides are assumed to be identical and with uniform nearest-neighbor coupling $\kappa_0$. In (b), the waveguides are identical while the coupling coefficients are subject to disorder, $\kappa_{ij} \sim \mathcal{U}(0.5\kappa_0,1.5\kappa_0)$. In (c), the coupling is uniform, $\kappa_0$, while the waveguides are detuned with propagation constants $\delta_{ij} \sim \mathcal{U}(-\delta_0 , \delta_0)$, where $\delta_0/\kappa_0 = 0.5$. In all three cases, the length is chosen such that $\kappa_0 L = 100$.}
    \label{fig3}
\end{figure}

It is straightforward to show that a random projector that respects the JL lemma cannot be a sparse matrix. This can be shown with a simple counterexample involving two orthogonal delta vectors, e.g., $\mathbf{x}_1=\delta_{l,l_1}$ and $\mathbf{x}_2=\delta_{l,l_2}$, where, $l=1,\cdots,d$ and $l_1 \neq l_2$. Clearly, a sparse random projector could map one of these two vectors to $\mathbf{0}$ and thus resulting in a large distance between the transformed vectors. Thus, one can say that a necessary condition for a JL projector is to be a dense matrix. According to this argument, the length of the photonic lattice should be large enough to guarantee the spreading of light from a single channel excitation (impulse response) across the entire array.

To understand the impact of transverse disorder in creating a proper random projector, the matrix elements of the complete evolution matrix $U$ is shown in Figs.~\ref{fig3} for three different cases of ordered and disordered lattices with off-diagonal and diagonal disorder. Here, the amplitude and phases of the matrix elements $u_{ij}$ and the corresponding histograms are depicted. According to this figure, for the first two cases the amplitudes of the matrix elements are centered around zero and the phases take discrete values of $0, \pm \pi/2, \pm \pi$. On the other hand, the matrix elements obey a completely different statistics in case of diagonal disorder. In this case, the amplitudes follow a Rayleigh distribution while the phases are uniformly distributed, or equivalently the real and imaginary parts of the matrix elements follow normal distributions.

\begin{figure*}[t]
    \centering
    \includegraphics[width=0.75\linewidth]{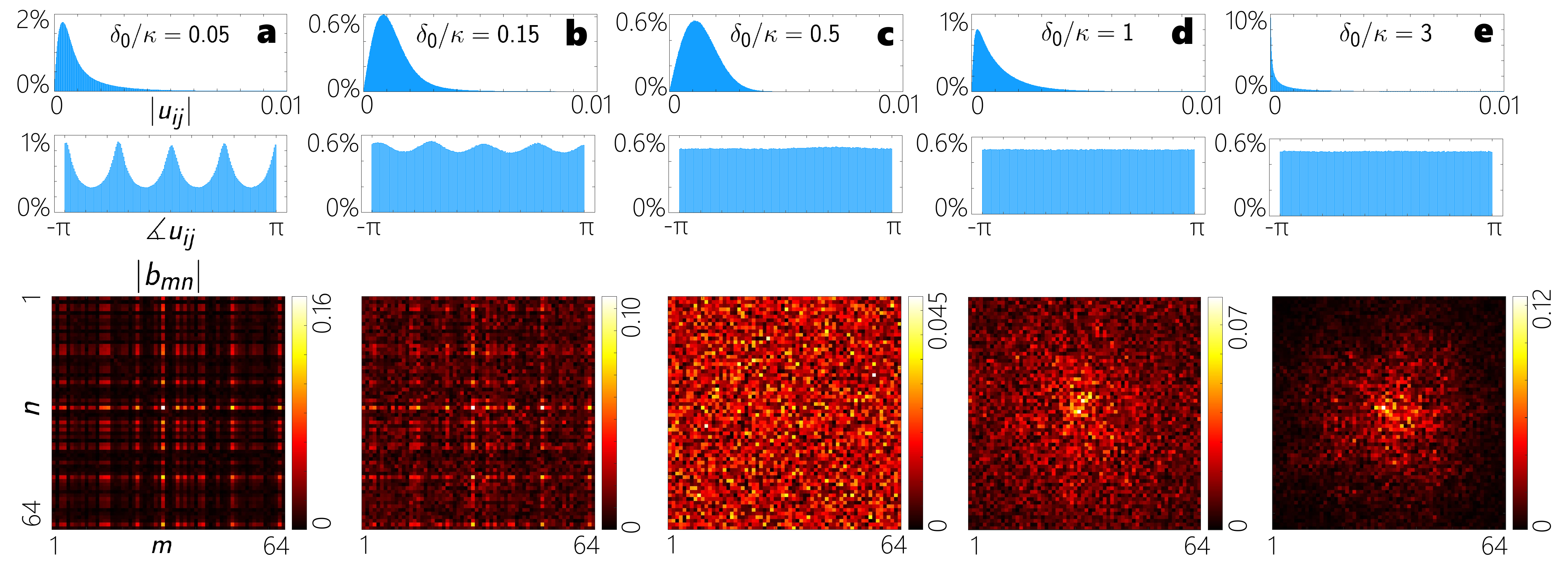}
    \caption{(a-e) The amplitude (top) and phase (middle) distributions of the evolution matrix elements $u_{ij}$ for different strengths of the diagonal disorder associated with the weak (a,b), intermediate (c) and strong (d,e) disorder regimes. The bottom panels show the associated output amplitude patterns when the central channel is excited (impulse response). Here a $d = 64 \times 64 $ lattice is considered. In all cases, uniform nearest-neighbor coupling $\kappa_0$ is considered such that $\kappa_0 L = 100$, where $L$ is the length of the array, and waveguide detunings are drawn from a uniform distribution $\delta_i \sim \mathcal{U}(-\delta_0, +\delta_0)$, where $\delta_0 / \kappa = 0.05$ (a), $0.15$ (b), $0.5$ (c), 1.0 (d) and $3$ (e).}
    \label{fig4}
\end{figure*}

The strength of the diagonal disorder is another important factor that plays a critical role in proper optical implementation of random projection. In fact, while a sufficient level of disorder is required to create the desired complex normal projector, strong disorder initiates transverse localization which is an undesired effect as discussed before. The amplitude and phase distribution of the matrix elements is plotted in Fig.~\ref{fig4} for different levels of disorder associated with ballistic transport, diffusive transport and transverse localization regimes. In addition, for a single channel excitation the intensity of light at the output plane is plotted for the three cases. This figure clearly indicates that the formation of the complex normal distribution requires a particular level of diagonal disorder. 

\begin{figure}[t]
    \centering
    \includegraphics[width=0.9\linewidth]{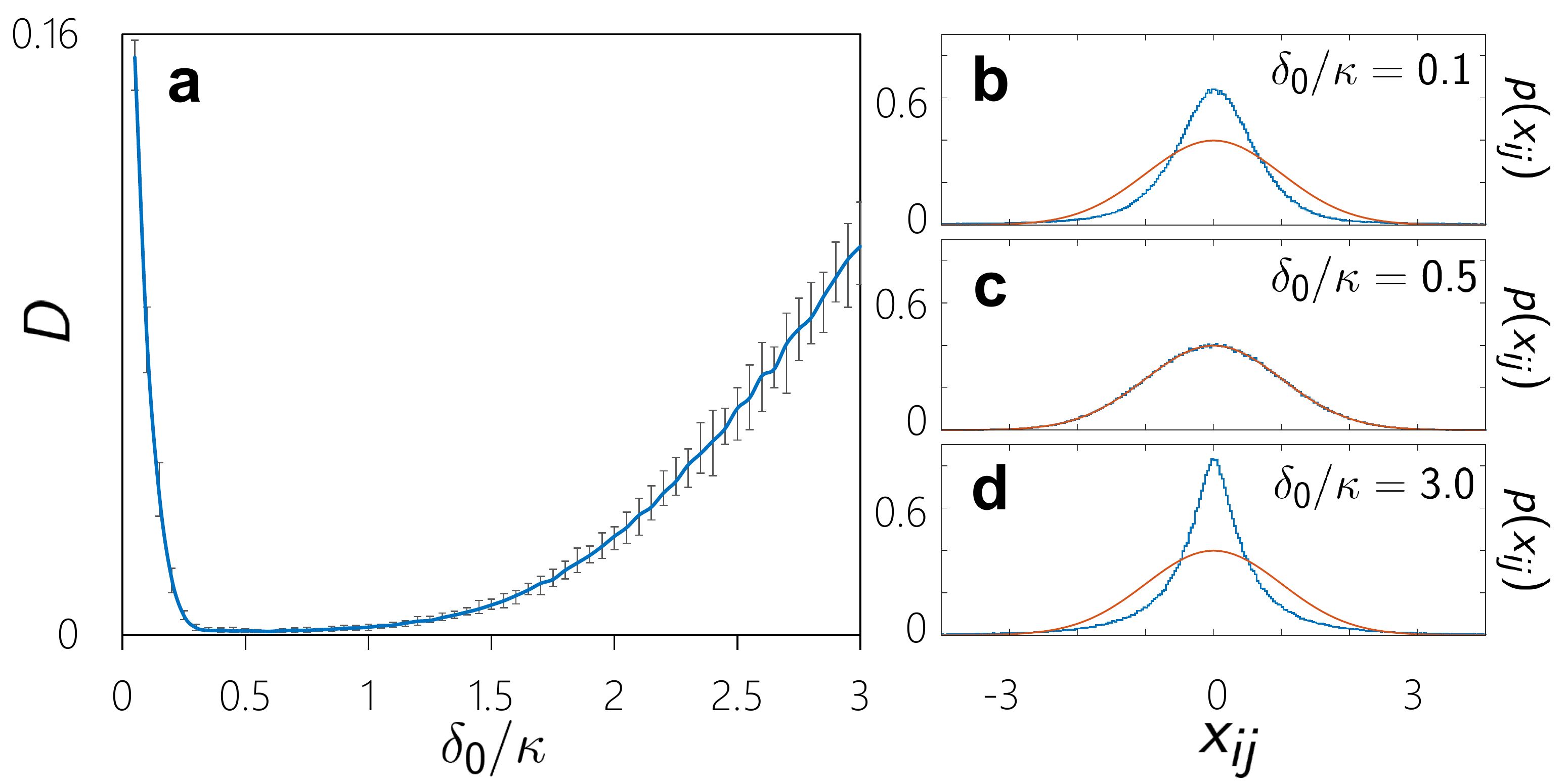}
    \caption{(a) Quantifying the level of diagonal disorder for random projection with a 2D waveguide array through the Kolmogorov--Smirnov statistic $D$. Here, each point represent the average of $D$ for 20 samples and the error bars mark the smallest and largest values. (b-c) Three sample distributions of the real parts of the matrix elements for weak ($\delta_0 = 0.1$), intermediate ($\delta_0 = 0.5$), and strong ($\delta_0 = 3$) disorder regimes. Here, an array of $d=32 \times 32$ waveguides with uniform coupling $\kappa_0$ are considered, while $\kappa_0 L =100$. In all cases, for comparison the sample distributions are normalized to have a standard deviation of unity.}
    \label{fig5}
\end{figure}

The latter observation could be better quantified through a statistical distribution test of entries of sample evolution matrices for different disorder levels. This can be done, for example, through a two-dimensional Kolmogorov--Smirnov test to evaluate the compatibility of the sample distribution of the matrix elements against a complex normal distribution \cite{fasano1987multidimensional}. Here, for simplicity, the real part of the matrix elements are considered and the Kolmogorov--Smirnov statistic $D$ is plotted for a range of the diagonal disorder levels in Fig.~\ref{fig5}. This statistic represents the largest deviation of the sample's empirical distribution function from the cumulative distribution function of a normal distribution. For illustration, the distributions of the real parts of the matrix elements are plotted for three samples in the weak, intermediate and strong disorder regimes. For the weak and strong disorder cases the tail of the sample distributions are located above the tail of the reference normal distribution. 

In conclusion, optical implementation of random projection by propagating spatially-modulated light through photonic waveguide arrays was suggested. The simplicity of the underlying principle allows for realistic implementation of optical random projection in different schemes. In particular, a photonic integrated implementation of a random projector can be realized through one-dimensional waveguide arrays. In addition, disordered multimode fibers \cite{mafi2019disordered} can be considered for two-dimensional fiber-based realizations. The proposed scheme can be utilized as a simple but powerful pre-processing stage for optical data processing and neural computing. In addition, optical random projection can have potential use in assisting conventional digital electronic computing when dealing with datasets of extremely-high-dimensions.

\begin{backmatter}

\bmsection{Acknowledgments}

The author gratefully thanks A. Genack and A. K. Jahromi for helpful discussions.

\bmsection{Disclosures}
The author declares no conflicts of interest.

\bmsection{Data Availability} Data underlying the results presented in Fig.~2 are available in Ref. \cite{ImageNetLink}. Other data can be obtained from the author.

\end{backmatter}



\bibliography{RP_Miri}

\begin{thebibliography}{10}
\newcommand{\enquote}[1]{``#1''}

\bibitem{wetzstein2020inference}
G.~Wetzstein, A.~Ozcan, S.~Gigan, S.~Fan, D.~Englund, M.~Solja{\v{c}}i{\'c},
  C.~Denz, D.~A. Miller, and D.~Psaltis, {\protect\JournalTitle{Nature}}
  \textbf{588}, 39 (2020).

\bibitem{bingham2001random}
E.~Bingham and H.~Mannila, \enquote{Random projection in dimensionality
  reduction: applications to image and text data,} in \emph{Proceedings of the
  seventh ACM SIGKDD international conference on Knowledge discovery and data
  mining,}  (2001), pp. 245--250.

\bibitem{liutkus2014imaging}
A.~Liutkus, D.~Martina, S.~Popoff, G.~Chardon, O.~Katz, G.~Lerosey, S.~Gigan,
  L.~Daudet, and I.~Carron, {\protect\JournalTitle{Scientific reports}}
  \textbf{4}, 1 (2014).

\bibitem{saade2016random}
A.~Saade, F.~Caltagirone, I.~Carron, L.~Daudet, A.~Dr{\'e}meau, S.~Gigan, and
  F.~Krzakala, \enquote{Random projections through multiple optical scattering:
  Approximating kernels at the speed of light,} in \emph{2016 IEEE
  International Conference on Acoustics, Speech and Signal Processing
  (ICASSP),}  (IEEE, 2016), pp. 6215--6219.

\bibitem{vellekoop2015feedback}
I.~M. Vellekoop, {\protect\JournalTitle{Optics express}} \textbf{23}, 12189
  (2015).

\bibitem{Gigan_2017}
S.~Rotter and S.~Gigan, {\protect\JournalTitle{Reviews of Modern Physics}}
  \textbf{89}, 015005 (2017).

\bibitem{goetschy2013filtering}
A.~Goetschy and A.~Stone, {\protect\JournalTitle{Physical review letters}}
  \textbf{111}, 063901 (2013).

\bibitem{davy2015transmission}
M.~Davy, Z.~Shi, J.~Wang, X.~Cheng, and A.~Z. Genack,
  {\protect\JournalTitle{Physical review letters}} \textbf{114}, 033901 (2015).

\bibitem{ImageNetLink}
\url{http://www.image-net.org/small/download.php}.

\bibitem{johnson1984extensions}
W.~B. Johnson and J.~Lindenstrauss, {\protect\JournalTitle{Contemporary
  mathematics}} \textbf{26}, 1 (1984).

\bibitem{frankl1988johnson}
P.~Frankl and H.~Maehara, {\protect\JournalTitle{Journal of Combinatorial
  Theory, Series B}} \textbf{44}, 355 (1988).

\bibitem{matouvsek2008variants}
J.~Matou{\v{s}}ek, {\protect\JournalTitle{Random Structures \& Algorithms}}
  \textbf{33}, 142 (2008).

\bibitem{achlioptas2003database}
D.~Achlioptas, {\protect\JournalTitle{Journal of computer and System Sciences}}
  \textbf{66}, 671 (2003).

\bibitem{ailon2009fast}
N.~Ailon and E.~Liberty, {\protect\JournalTitle{Discrete \& Computational
  Geometry}} \textbf{42}, 615 (2009).

\bibitem{segev2013anderson}
M.~Segev, Y.~Silberberg, and D.~N. Christodoulides,
  {\protect\JournalTitle{Nature Photonics}} \textbf{7}, 197 (2013).

\bibitem{lahini2008anderson}
Y.~Lahini, A.~Avidan, F.~Pozzi, M.~Sorel, R.~Morandotti, D.~N. Christodoulides,
  and Y.~Silberberg, {\protect\JournalTitle{Physical Review Letters}}
  \textbf{100}, 013906 (2008).

\bibitem{schwartz2007transport}
T.~Schwartz, G.~Bartal, S.~Fishman, and M.~Segev,
  {\protect\JournalTitle{Nature}} \textbf{446}, 52 (2007).

\bibitem{di2013einstein}
G.~Di~Giuseppe, L.~Martin, A.~Perez-Leija, R.~Keil, F.~Dreisow, S.~Nolte,
  A.~Szameit, A.~Abouraddy, D.~Christodoulides, and B.~Saleh,
  {\protect\JournalTitle{Physical review letters}} p. 150503 (2013).

\bibitem{christodoulides2003discretizing}
D.~N. Christodoulides, F.~Lederer, and Y.~Silberberg,
  {\protect\JournalTitle{Nature}} \textbf{424}, 817 (2003).

\bibitem{fasano1987multidimensional}
G.~Fasano and A.~Franceschini, {\protect\JournalTitle{Monthly Notices of the
  Royal Astronomical Society}} \textbf{225}, 155 (1987).

\bibitem{mafi2019disordered}
A.~Mafi, J.~Ballato, K.~W. Koch, and A.~Sch{\"u}lzgen,
  {\protect\JournalTitle{Journal of Lightwave Technology}} \textbf{37}, 5652
  (2019).

\end{thebibliography}


\end{document}